\documentclass[aps,prl,amsmath,amssymb,superscriptaddress]{revtex4}
\usepackage{graphicx}
\usepackage{dcolumn}
\usepackage{color}
\usepackage{bm}
\usepackage{amsmath,amssymb}
\usepackage[normalem]{ulem}

\begin{document}



\title{Experimental realization of a topological p--n junction \\by intrinsic defect-grading }

\author{T.\,Bathon}
\affiliation{Physikalisches Institut, Experimentelle Physik II,
	Universit\"{a}t W\"{u}rzburg, Am Hubland, D-97074 W\"{u}rzburg, Germany}
\author{S.\,Achilli}
\affiliation{Fisica, Universit\`{a} Cattolica di Brescia, via dei Musei 41, I-25121 Brescia, Italy}
\author{P.\,Sessi}
\email[corresponding author: ]{sessi@physik.uni-wuerzburg.de}
\affiliation{Physikalisches Institut, Experimentelle Physik II,
	Universit\"{a}t W\"{u}rzburg, Am Hubland, D-97074 W\"{u}rzburg, Germany}
\author{V.~A. Golyashov}
 \affiliation{A.V. Rzanov Institute of Semiconductor Physics, Siberian Branch, 
 	Russian Academy of Sciences, 630090 Novosibirsk,  Russia}
\author{K.\,A.\,Kokh}
\affiliation{V.S. Sobolev Institute of Geology and Mineralogy, Siberian Branch, 
	Russian Academy of Sciences, 630090 Novosibirsk,  Russia}
\affiliation{Novosibirsk State University, 630090 Novosibirsk,  Russia}
\affiliation{Saint-Petersburg State University, 198504 Saint-Petersburg, Russia}
\author{O.\,E.\,Tereshchenko}
\affiliation{A.V. Rzanov Institute of Semiconductor Physics, Siberian Branch, 
	Russian Academy of Sciences, 630090 Novosibirsk,  Russia}
\affiliation{Novosibirsk State University, 630090 Novosibirsk,  Russia}
\affiliation{Saint-Petersburg State University, 198504 Saint-Petersburg, Russia}
\author{M.\,Bode}
\affiliation{Physikalisches Institut, Experimentelle Physik II,
	Universit\"{a}t W\"{u}rzburg, Am Hubland, D-97074 W\"{u}rzburg, Germany}
\affiliation{Wilhelm Conrad R{\"o}ntgen-Center for Complex Material Systems (RCCM),
	Universit\"{a}t W\"{u}rzburg, Am Hubland, D-97074 W\"{u}rzburg, Germany}



\date{\today}
\vspace{1cm}
\begin{abstract}
\vspace{1cm}
\bf{A junction between an n- and p-type semiconductor results in the creation of a depletion region whose properties are at the basis of nowadays electronics. If realized using topological insulators as constituent materials, p-n junctions are expected to manifest several unconventional effects with great potential for applications. Experimentally, all these fascinating properties remained unexplored so far, mainly because prototypical topological PNJs, which can be easily realized and investigated, were not readily available.
Here, we report on the creation of topological PNJs which can be as narrow as few tenths of nm showing a built-in potential of 110meV. These junctions are intrinsically obtained by a thermodynamic control of the defects distribution across the crystal. Our results make Bi2Te3 a robust and reliable platform to explore the physics of topological p-n junction.}
\vspace{1cm}
\noindent
\end{abstract}

\pacs{}
\keywords{}
\maketitle



The recent discovery of topologically protected surface states in the binary chalchogenides Bi$_2$Te$_3$ \cite{CAC2009}, Bi$_2$Se$_3$ \cite{XQH2009} and Sb$_2$Te$_3$\cite{JWC2012} has tremendously revitalized the interest in these narrow gap semiconductors that have been studied for decades because of their excellent thermoelectric properties \cite{VSC2001}. Topological insulators (TIs) are materials insulating in the bulk but metallic on their surface due to the existence of linearly dispersing gapless states which, contrary to the trivial surface states usually found in metals and semiconductors, are protected by time reversal symmetry \cite{HK2010}. The strong spin-orbit coupling characterizing these materials perpendicularly locks the spin to the momentum, resulting in a chiral spin texture which forbids backscattering \cite{RSP2009,ZCC2009} and makes spin currents intrinsically related to charge currents \cite{KVB2007}.

These unconventional properties make TIs an ideal platform to realize exotic states of matter, such as Majorana fermions \cite{FK2008} and magnetic monopoles \cite{QLZ2009}. Furthermore, they appear to be ideal candidates to realize magneto-electric and spintronics devices with low power consumption \cite{GF2010}. Within this framework, the creation of p--n junctions (PNJ), which are the building blocks of several semiconducting devices, such as diodes, sensors, solar cells, or transistors, in topological materials would represent the first step towards the direct application of this fascinating class of materials \cite{WCZ2012}. In TIs, the massless character of surface carriers and the helical spin texture inversion at the p--n interface are predicted to result in several unconventional phenomena, e.g. gapless chiral edge mode \cite{WCZ2012}, Klein tunneling \cite{KNG2006}, Veselago lenses \cite{CFA2007}, exciton condensation and charge fractionalization \cite{SMF2009}.

Experimentally, all these fascinating properties remained unexplored so far, mainly because prototypical topological PNJs, which can be easily realized and investigated, were not readily available. One issue towards the successful fabrication of topological PNJs is the presence of defects which are intrinsically incorporated in TI materials during the growth process. Thereby, the resulting TI crystals are heavily doped at a relatively high bulk carrier concentration level \cite{SKS2012,WHT2013}. Since bulk carriers give rise to leakage currents that make capacitive charging impossible their existence represent a severe obstacle towards the creation of topological PNJs by applying a gate bias between opposite surfaces of a TI thin film \cite{SMF2009}.

Recent studies showed, however, that---depending on the growth conditions---large deviations from the nominal composition can be obtained \cite{MBB2013}. These defects, which can be anti-sites or vacancies, are characterized by different formation energies. Here, we demonstrate that their inevitable presence can be turned into a positive effect by thermodynamically controlling their distribution across the crystal. In particular, we show how appropriate growth conditions result in the creation of bulk crystals that intrinsically contain topological PNJs at their surface, thereby avoiding the complicated fabrication of heterostructures and the problems related to the creation of interfaces between different materials \cite{ZMF2013,ZCJ2013,ZHX2014}. Atomic scale scanning tunneling microscopy (STM) and spectroscopy (STS) measurements combined with {\it ab initio} calculations evidence that Te segregation \cite{WZS2011,JBD2012,WHT2013} results in well-separated Te- and Bi-rich regions that display p- and n-transport character, respectively. Spatially resolved Hall and Seebeck measurements confirm the transition from p- to n-like transport. At the p--n interface the oppositely charged defects, i.e.\ donors or acceptors, compensate, resulting in a substantial drop of conductivity by almost two orders of magnitude. Scanning tunneling spectroscopy reveals a built-in potential of about 110\,meV, thereby considerably shifting the Dirac point in between the p- and n-region. Spectroscopic data obtained within the p--n compensation region indicate that the junction width amounts to several 10\,nm only.  

\begin{figure}[t]   
\includegraphics[width=.6\columnwidth]{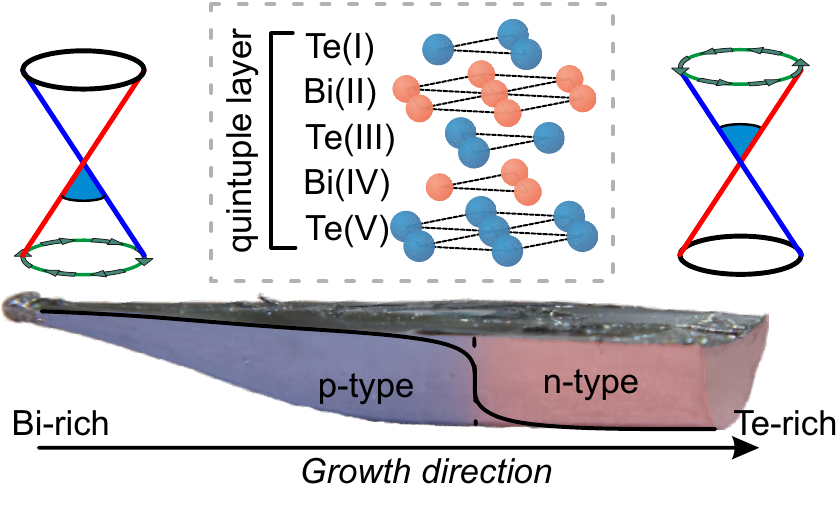}%
\caption{Lower panel: Photographic image of a Bi$_2$Te$_3$ crystal grown by the modified Bridgman technique. As schematically illustrated by different colors a doping gradient leads to a position-dependent transition from p- to n-type conductivity with the chemical potential lying in the upper and lower part of the Dirac cone, respectively. This transition is accompanied by an inversion of the helical spin texture, as indicated by green arrows in the upper left and right panel. The crystal structure is shown in the hatched box.}
\label{fig:crystal}
\end{figure}  

Figure\,\ref{fig:crystal} schematically presents some essential background information related to our approach. Within the hatched box of Fig.\,\ref{fig:crystal} the crystal structure of a quintuple layer of the binary chalchogenide Bi$_2$Te$_3$ is shown. It consists of alternating Te and Bi layers. Adjacent quintuple layers are weakly bound by van der Waals forces thereby offering a natural cleavage plane, a situation highly advantageous for STM experiments. The bottom panel of Fig.\,\ref{fig:crystal} shows a photographic image of a Bi$_2$Te$_3$ crystal grown by the modified Bridgman technique (see Experimental Section for details). The crystal has a nominal composition with a Te content of 61 mol\%. As recently shown, the conduction of Bi$_2$Te$_3$ crystals changes from p- to n-type at approximately this Te concentration \cite{KMG2014}. Due to Te segregation, however, the stoichiometry will not be constant but continuously change during the growth process from Bi-rich to Te-rich, as indicated by different color shadings in the bottom panel of Fig.\,\ref{fig:crystal}. Correspondingly, we expect Dirac points which are energetically located below and above the Fermi level, respectively, with opposite rotational sense of the spin polarization. As we move from p- to n-doped surface areas we expect to pass through a transition region with equal concentrations of Bi- and Te-induced charge carriers where the carrier concentration becomes minimal, thereby realizing an intrinsic TI.


\begin{figure}[t]  
\includegraphics[width=0.95\columnwidth]{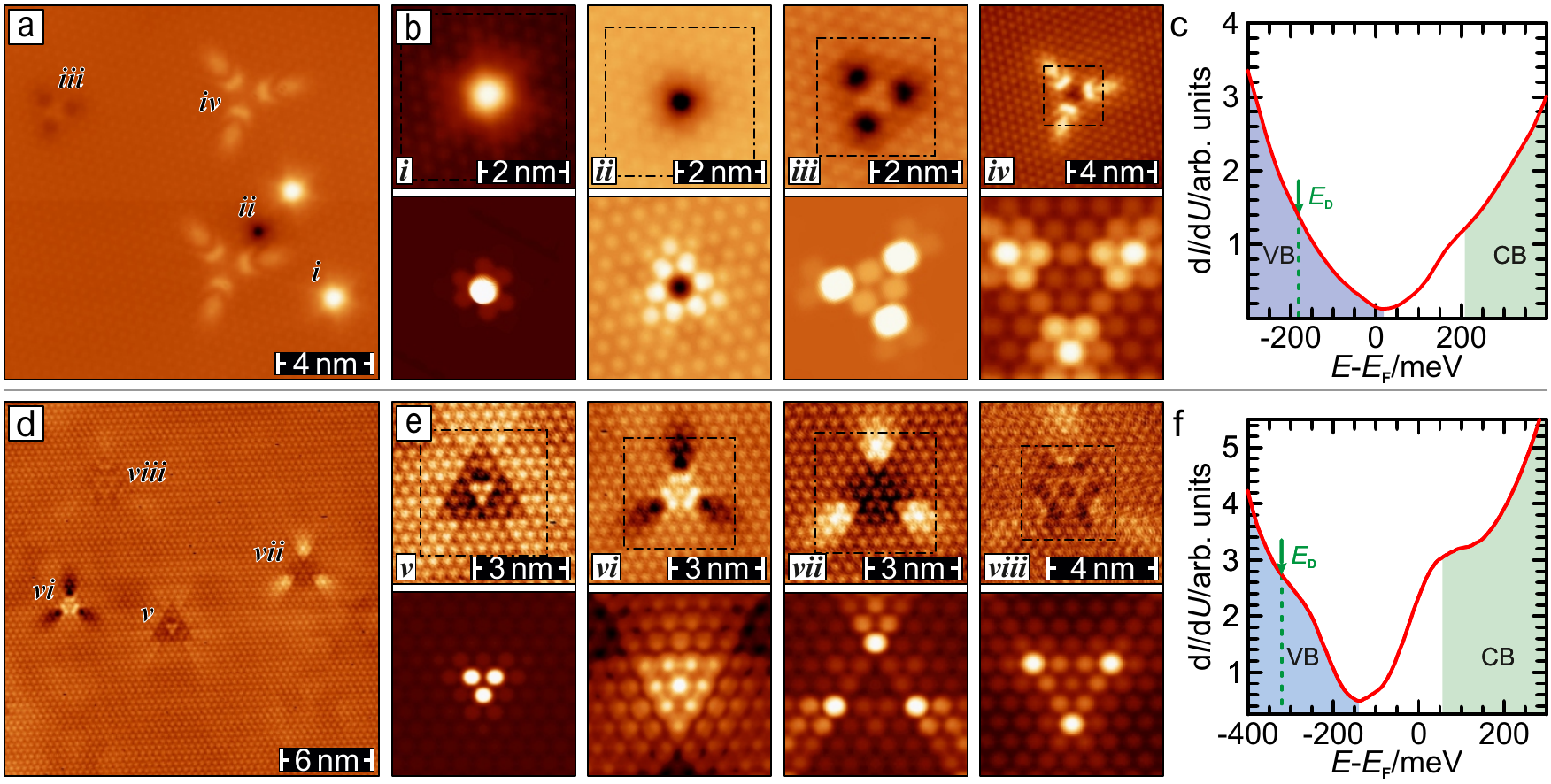}%
\caption{(a) Constant-current image obtained over a Bi-rich crystal region. Four different defects (i)-(iv) can be recognized. They are imaged on the atomic scale by STM in the top row of (b) and compared to {\em ab-initio} calculations which, due to the elevated computational cost required by the large number of atoms, have been restricted over the areas identified by the hatched boxes visible in the upper frames. This comparison between theory and experiment allows for their unambiguous identification (see text for details). (c) STS data evidence a Fermi level in close proximity of the top of the valence band (VB), an observation which allows to identify this crystal region as p-doped. $E_{\rm D}$ indicates the Dirac energy of the surface state. (d) Constant-current image obtained over a Te-rich crystal region with four characteristic different defects labeled (v)-(viii). (e) Again comparison of atomic scale images with {\em ab-initio} calculations allows to unequivocally position them in a well-defined atomic plane of the quintuple layer structure. (f) STS data identify this crystal region as n-doped due to the proximity of the Fermi level with the conduction band (CB).}
\label{fig:STM}
\end{figure}  

In fact, the differently doped areas can clearly be distinguished by STM and STS. Fig.\,\ref{fig:STM}(a) displays an overview topographic STM image obtained at the (0001) surface of the Bi-rich crystal region. Four different defects labeled (i)-(iv) are visible on the surface. Since the defects symmetry and extension reflects the perturbation introduced by the bonding structure within the crystal \cite{PhysRevLett.108.066809} (see bond geometry sketched in the inset of Fig.\,\ref{fig:crystal}), a detailed analysis of the atomically resolved images reported in the top row of Fig.\,\ref{fig:STM}(b) allows to identify their location within the quintuple layer structure of Bi$_2$Te$_3$, i.e.\ Te(I)-Bi(II)-Te(III)-Bi(IV)-Te(V), where Te(I,V) and Te(III) represent two inequivalent Te planes. Following this procedure the defects (i)-(iv) are all located within Te layers, and they can be Te vacancies (V$_{\rm Te}$) or antisites (Bi$_{\rm Te}$). Due to the lack of chemical sensitivity of STM a definite assignment can only be achieved by a comparison of the experimental data with simulated STM images obtained from {\itshape ab initio} calculations [bottom row of Fig.\,\ref{fig:STM}(b)]. Based on the good general agreement achieved between the experimental and theoretical images we can make the following defect assignment: defect (i) is a Bi$_{{\rm Te(I)}}$ antisite; defect (ii) is a Te surface vacancy (V$_{\rm Te(I)}$); defect (iii) is a Te vacancy in the third layer (V$_{\rm Te(III)}$); and defect (iv) is a Bi antisite in the fifth layer (Bi$_{\rm Te(V)}$). Since all defects are located in Te layers this region is expected to be Te-poor. This is confirmed also by theoretical calculations which show that these defects are characterized by the lowest formation energy in Bi-rich regions \cite{SKS2012}. The electronic properties of this sample region can be characterized by an investigation of the local density of states as inferred by the STS spectra reported in Fig.\,\ref{fig:STM}(c). Following the energy level positioning scheme adopted in Ref.\,\cite{SOB2013} we obtain a valence band maximum close to the Fermi level, thereby indicating the p-doped character of this sample region.

Investigation of the opposite side of the crystal reveals four other defects (v)-(viii) [Fig.\,\ref{fig:STM}(d)], the STM appearance of which is quite different from the data presented in Fig.\,\ref{fig:STM}(a)-(b) before. Again the symmetry and lateral extension of the various defects in atomic resolution data were used to estimate the position of the defects. Based on these estimations {\em ab initio} DFT calculations were performed. The comparison presented in Fig.\,\ref{fig:STM}(e) reveals a very reasonable agreement. Our results show that all four defects are located in the two Bi layers of the Bi$_2$Te$_3$ quintuple layer, in agreement with the expected Te-rich stoichiometry. Namely, the following assignments were concluded: defect (v) is a Te$_{\rm Bi(II)}$ antisite; defect (vi) is a  Bi vacancy (V$_{\rm Bi(II)}$); defect (vii) is a Te$_{\rm Bi(IV)}$ antisite; and defect (viii) is a Bi vacancy, V$_{\rm Bi(IV)}$. As for the p-doped region, our defects identification is consistent with the formation energies calculated by Scanlon {\em et al.}\,\cite{SKS2012}, which show that in Bi$_2$Te$_3$ grown under Te-rich conditions the lowest energy defects are Te$_{\rm Bi}$ antisites and Bi vacancies, with the last ones playing a much less significant role. Comparison of the STS spectra shown in Fig.\,\ref{fig:STM}, panels (c) and (f), reveals that the different types of defects present in Bi- and Te- rich areas strongly influence the respective electronic properties, i.e.\ Te-rich regions show a rigid spectral shift towards negative energies with respect to Bi-rich areas. This negative shifts amounts to approximately 140\,meV between the p- and n-doped surface areas of the Te-rich and Bi-rich parts of the crystal, respectively. This observation implies that a topological PNJ is naturally present within our crystals.


\begin{figure}[t]   
\begin{minipage}[t]{0.55\textwidth}
\includegraphics[width=.75\textwidth]{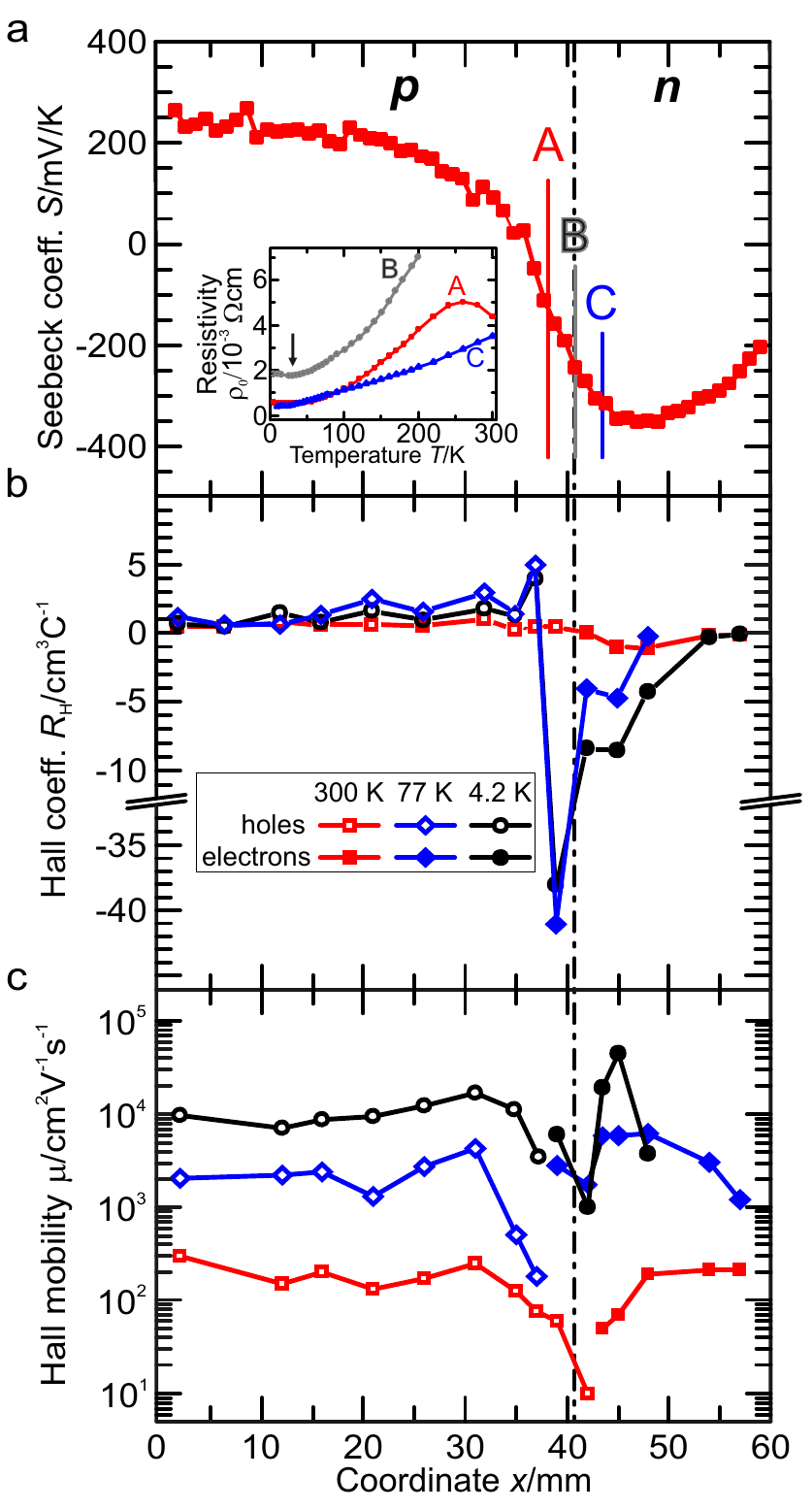}%
\end{minipage}
\hfill
\caption{(a) Room temperature Seebeck coefficient $S$ measured along the crystal growth direction $x$ with indication of three points (A,B,C) in the p--n transition region where electrical resistivity $\rho_0$ (inset) was measured. (b) Hall coefficient $R_\text{H}$ measured at three temperatures $T$. Current and the magnetic field direction was within and perpendicular to the cleavage plane, respectively. (c) Hall mobility $\mu$ with respect to the position $x$ along the boule.} \label{fig:Transport}
\end{figure}   

Figure \ref{fig:Transport}(a) shows the room-temperature Seebeck coefficient $S$ of a Bi$_2$Te$_3$ single crystal measured at various positions (0--60\,mm) along the crystal growth direction $x$. It continuously and slowly changes along the crystal rod with a sign change occurring at a position of 35\,mm. A positive Seebeck coefficient in the range of 250--100\,$\mu$V/K is obtained for the left part of the crystal boule (starting from $x=0$). Further right to this region, i.e.\ between 30\,mm and about 35\,mm, the Seebeck coefficient rapidly drops and eventually becomes negative, indicating that the crystal conductivity changes from p- to n-type at the point where $S \sim 0\,\mu$V/K. In the region of the p--n transition ($S \sim 0\,\mu$V/K) a potential Seebeck microprobe demonstrates the change of sign of the Seebeck coefficient on the 200 $\mu$m distance, limited by the resolution of the method. At $x\approx 47$\,mm a high value $S = -390\,\mu$V/K is obtained.

This interpretation is also corroborated by Hall measurements presented in Fig.\,\ref{fig:Transport}(b) which were taken at the bottom and top parts of the boule. They show opposite signs of Hall coefficients $R_{\rm H}$, thereby confirming the presence of both p- and n-type regions. The left part of the boule (0--30\,mm) shows metallic behavior with small positive Hall coefficients leading to a weakly temperature-dependent p-type carrier concentration, $p = (5-10) \times 10 ^{18}$ cm$^{-3}$. At the right part a similar metallic-like behavior was observed for n-type carriers ($n \simeq 10 ^{19}$cm$^{-3}$). The large carrier concentration and its weak temperature dependence indicate that the bottom and top parts of the boule are heavily doped by acceptors and donors, respectively. Due to a strong reduction of the carrier concentration by two orders of magnitude ($\sim 10^{17}$\,cm$^{-3}$) in the vicinity of the p--n junction, the Hall coefficient changes sign and drops down to $-45$\,cm$^{3}$C$^{-1}$ at low temperatures. These data indicate that an intrinsic binary chalchogenide TI can be obtained by a minimization of the carrier concentration through a careful control of the alloy composition between Bi and Te.

Electrical resistivity temperature dependencies were measured on the samples taken from positions A, B, and C [cf.\ Fig\,\ref{fig:Transport}(a)], i.e.\ within a distance of 2\,mm of the sign change of the Hall coefficient. In all the points resistivity decreases for temperatures from RT to $T\approx 30$\,K, and then remains almost constant down to the liquid helium temperatures for p- and n- type samples, while for point B resistivity increases, demonstrating semiconducting behavior. In both p- and n-parts of the crystal, the mobilities show a high average value of $10^{4}$\,cm$^{2}$V$^{-1}$s$^{-1}$ at low temperatures [Fig\,\ref{fig:Transport}(c)]. At $x=45$\,mm in the n-type region, i.e.\ close to the p--n transition, $\mu$ reaches the maximum value of about $10^{5}$\,cm$^{2}$V$^{-1}$s$^{-1}$. At the same position a maximum Seebeck coefficient $S \sim -400\,\mu$V/K is observed, meaning that near the p-n transition there is the suppression of the defects formation.


\begin{figure}[t]   
\includegraphics[width=0.95\columnwidth]{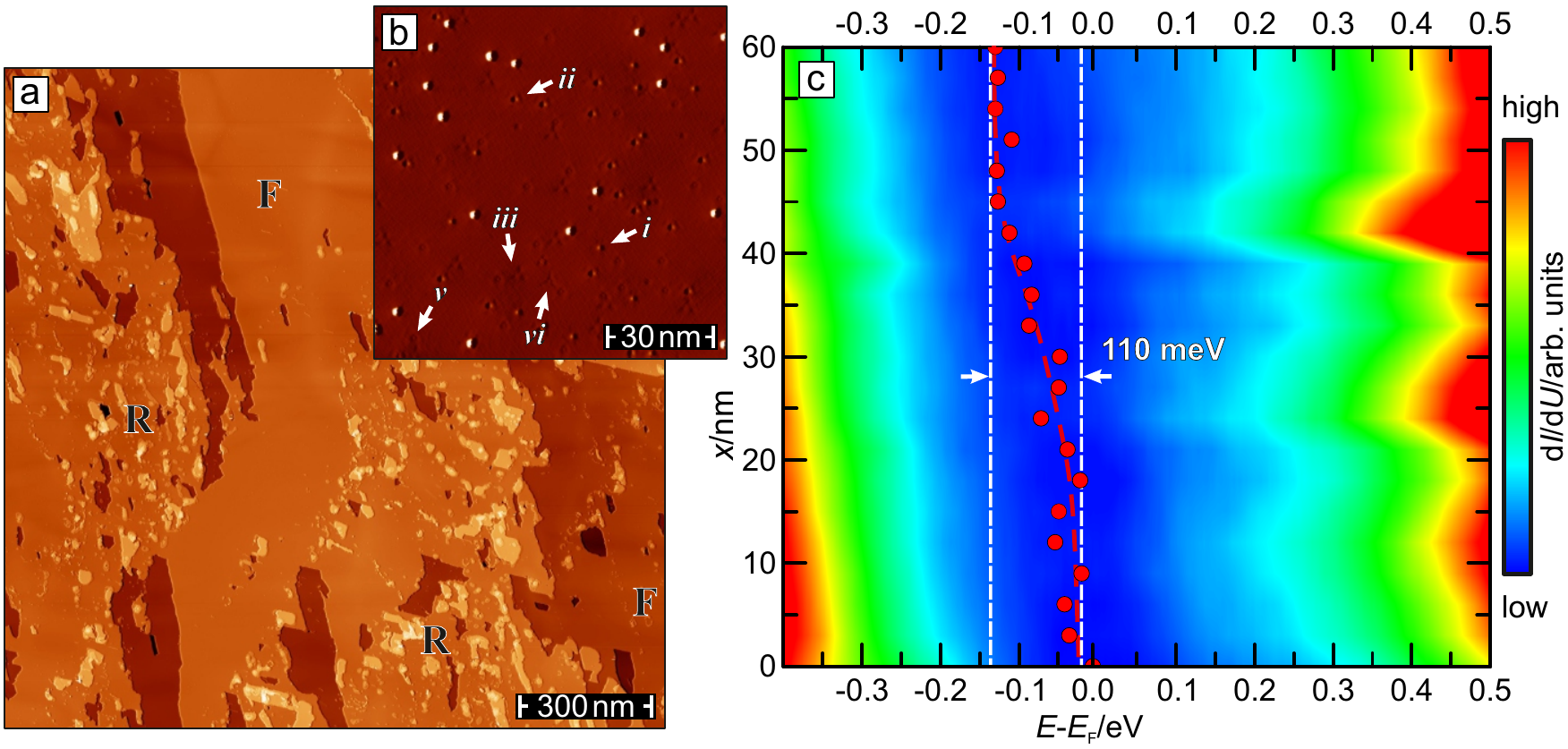}
\caption{(a) Overview topographic STM image of the Bi$_2$Te$_3$ crystal taken in the p--n transition zone (scan range: 1.5\,$\mu$m\,$\times$\,1.5\,$\mu$m). The surface shows two characteristic regions each of which is several hundred nm wide; while the rough surface region (R) contains numerous islands and clusters the other, the flat region, exhibits atomically flat terraces (F) that are separated by unit cell step edges. The zoomed in image (b) was taken within a flat region. Defects characteristic for both, n- (i-iv) and p-type (v-viii) Bi$_2$Te$_3$ can be found. (c) Sequence of color-coded tunneling spectra taken along a 60\,nm long line (3\,nm increment) within a flat region. The energetic position of the minimum, which is marked by a red point in each spectrum, shifts from just above the Fermi level (bottom) to about 110\,meV, indicative for a p--n transition with a width of about 40\,nm. The hatched lines are guides to the eye.}
\label{fig:p--n}
\end{figure}  

In order to investigate the minimal width of intrinsic TI p--n transitions that result from the slightly graded stoichiometry during the crystallization process, we have performed STM/STS in the zone where the transition was detected in the transport measurements described above. An overview scan is displayed in Fig.\,\ref{fig:p--n}(a). While the surface region in the left and the bottom right part of the image is relatively rough (marked R) because of numerous small islands and clusters, we also find regions with extended atomically flat terraces separated by unit-cell step edges (F).  As we zoom into such an atomically flat region [Fig.\,\ref{fig:p--n}(b)] we can observe the coexistence of defects which we identify as being characteristic for n- and p-type Bi$_2$Te$_3$, i.e.\ defects (i), (ii), and (iii) as well as (v) and (vi), respectively (see above). This microscopic picture explains the drop of the Hall coefficient observed in transport, see Fig.\,\ref{fig:Transport} and Ref.\,\onlinecite{JBD2012,MBB2013}. It is also in agreement with the mobility drop detected in the p-n junction region, which can be thus be directly linked to the increased number of total defects. Overall, these findings have deep implications going beyond the scope of the present work. Minimizing bulk conductivity has always been considered as a powerful strategy to let topological surface states dominate the scene. However, it was not clear up to know if this effect should be ascribed to defects suppression or compensation. The combination of transport measurements with atomic scale characterization techniques provides thus a detailed microscopic picture solving this problem, identifying defect compensation as the responsible mechanism. Defect suppression can be safely ruled out since it would require a significant reduction of the defects concentration, a condition not consistent with our results.

Local spectroscopic data show that these regions contain several PNJs which are very sharp. This is corroborated by the color-coded STS data shown in Fig.\,\ref{fig:p--n}(c) measured with a 3\,nm increment along a 60\,nm long line in a flat region. The minimum position is marked by a red dot in each spectrum. Indeed we observe a shift of the minimum from close to the Fermi level $E_{\rm F}$ at $x \le 0$\,nm to $E - E_{\rm F} \approx 110$\,meV at $x \ge 50$\,nm, i.e.\ the transition from p- to n-doped Bi$_2$Te$_3$ takes place over the remarkably short distance of about 40\,nm, although the p- or n-type character observed at the two ends of the boule [cf. Fig.\,\ref{fig:STM}(c) and (f)] is not fully reached.

Experimental data obtained on a larger length scale indicate that the doping gradient is not monotonic but exhibits some fluctuations which lead to the coexistence of several p--n transition regions. Their distance typically amounts to $\approx 1\,\mu$m, a value also matching the length scale of flat (F) and rough (R) surface patches in Fig.\,\ref{fig:p--n}(a). Both behaviors potentially reflect the complicated convection processes that take place during the slow cooling and crystallization process of the boule. Their averaged random distribution result in a Ohmic characteristic preventing spatially averaging technique from verifying the existence of a rectifying behavior, an aspect which may be directly tackled by using nanoscale four probes techniques. Although their creation is not yet fully controlled, our data indicate that p--n transition regions naturally resulting from the slightly graded stoichiometry of crystals may be very narrow. Further research may lead to optimized procedures which better control their position and width to make them useful for application.

In summary, we demonstrated that, contrary to other materials, the ambipolar behavior of Bi$_2$Te$_3$ makes possible, by taking advantage of the intrinsic defects which are unavoidably introduced during the growth process, to create topological p--n junctions. Scanning tunneling spectroscopy reveals that their width amounts to 40\,nm only. The related potential drop of approximately 110\,meV shall allows for direct applications of PNJs in room temperature devices. Additionally, the defect compensation naturally achieved at the p--n interface results in a strong reduction of the bulk carriers concentration, thus paving the way to explore the PNJ topological properties.

\section{Acknowledgments}
The authors gratefully acknowledge stimulating discussions with Mario Italo Trioni.
This work was supported by the Deutsche Forschungsgemeinschaft within SPP 1666 (Grant No. BO1468/21-1). 
K.A.K. and O.E.T. acknowledge the financial support by the RFBR (Grant nos. 13-02-92105 and 14-08-31110).  

\section{Experimental section}

Bi$_2$Te$_3$ crystals have been grown by the modified Bridgman technique with a temperature gradient of about 10\,K/cm at the front of crystallization \cite{KMG2014}. As-grown ingots had a single crystalline structure and were split into two parts along the cleavage plane (0001) oriented along the growth direction (Fig.\,\ref{fig:crystal}). One part of each crystal was cut perpendicular to the growth axis into 0.5--1\,mm samples. Indium solder Ohmic contacts were used for transport measurements. The Hall resistance $R_{yx}$ and the resistance $R_{xx}$ were measured in the Hall bar geometry using a standard six-probe method on rectangular samples. A potential Seebeck microprobe was used to investigate the room-temperature Seebeck coefficient in the region of p--n transition with a spatial resolution of 200\,$\mu$m. The Hall mobility was calculated from the measured conductivity and calculated carrier concentration, which was extracted from the measured Hall coefficient: n=1/(Rh*e). The carrier concentration dependence along the crystal was shown in \cite{KMG2014}.

STM measurements were performed at a tip and sample temperature $T = 4.8$\,K with electro-chemically etched tungsten tips. After transfer into the ultra-high vacuum system, samples were cleaved at room temperature at a base pressure $p < 2 \times 10^{-11}$\,mbar and immediately inserted into the cryogenic STM. The PNJ was located by bringing the tip into tunneling distance from the surface and recording a local tunneling spectrum indicative for p- or n-type doping [cf. Fig.\,\ref{fig:STM}(c) and (f)]. After retracting the tip the sample was moved with an $x-y$-stage (initially by several tens to hundreds of $\mu$m). This procedure was repeated until a spectral shift signaled that the boundary to the region governed by the other dopant had been crossed. Then a refined procedure with smaller $x-y$-movements was performed.

\newpage
\section{Theoretical calculation}
{\em Ab-initio} theoretical calculations were performed by density functional theory, using a pseudo-potential representation of the electron--ion interaction and local orbital basis sets, as implemented in the SIESTA code \cite{SIESTA}. We used a generalized gradient approximation (GGA) for the exchange-correlation functional \cite{PBE} and a plane wave cutoff equal to 250\,Ry. The experimental lattice constant for the hexagonal cell of Bi$_2$Te$_3$ was used \cite{EXPLA} and a $8 \times 8$ surface cell was adopted to reduce the lateral interaction between crystal defects. The slab thickness was chosen in order to include three of the quintuple layers forming the building block of Bi$_2$Te$_3$ in the direction orthogonal to the surface and a large portion of vacuum of approximatively 30\,\AA. Constant--distance STM images have been simulated in the Tersoff-Hamann model \cite{TH} by calculating the Kohn-Sham local density of states (LDOS) in the energy interval between the Fermi level $E_{\rm F}$ and $E_{\rm F} + eV_{\rm b}$ \cite{STM}, where the applied bias $eV_{\rm b}$ was fixed in agreement with the experimental setup. In particular the moderately large value of $eV_b$, of the order of 0.4 eV, allows us to obtain numerically stable results and to reasonably neglect the small inaccuracy of the LDOS around the Fermi level due to the absence of spin-orbit interaction in the calculation. A tip--surface distance equal to 3\,{\AA} was considered, although we verified that the result of our STM simulations are robust against small variations of this value. The extension of the tip was taken into account in the STM simulation by considering the mean value of LDOS within a radius of 4\,{\AA} around the tip position. 

\bibliography{pn_References}

\end{document}